# The Diffusion-Hardening Effect on the Technological Properties of High-Temperature Steel

**A.V. Hruzevyich[1,2], \*D.O. Derecha[2,3]**

[1] Trypil'ska CHP, UA-08720 Ukrainka, Kyiv Region, Obukhov District, Ukraine

[2] Institute of Magnetism Nat. Acad. of Sci. of Ukraine and Min. of Edu. And Sci. of Ukraine, 36b Academician Vernadsky Blvd., UA-03142 Kyiv, Ukraine

[3] National Technical University of Ukraine "Igor Sikorsky Kyiv Polytechnic Institute" 37, Prosp. Peremohy, Kyiv, UA-03056, Ukraine

*e-mail:* dderecha@gmail.com

PACS numbers: 06.20.−f, 44.90.+c, 61.05.−a, 61.50.−f, 61.72.−y, 61.82.Bg

## Abstract

The results of research pipes diffusion hardening as an effective method for increasing the durability and reliability of power equipment are presented. The experience of commercial operation of pipes manufactured from diffusion chromized and heat-hardened steel DIN: 1.7715; WNr: 14 MoV6-3; BS: 1503-660-440 on the heating surfaces of the supercritical pressure boilers are generalized. The possibility and effectiveness of this method on the example of capacities of Trypil'ska CHP are shown.

Keywords: diffusion-hardening, steel, durability, structure, heat resistance, weldability, corrosion resistance, heat resistance.

# 1. Introduction

Modern metal materials which are used in power engineering, machine building, ship-building and construction industry should provide the necessary resource and reliability of the structures. The most important characteristics that determine the reliability of the work of components and constructions are the wear indexes, long-time strength and corrosion characteristics. The corrosive characteristics and characteristics of mechanical degradation are largely dependent on the condition of the components surface. The material surface condition significantly effects on the cracking rate in the material, therefore, for the improvement of the metal components characteristics the matter of technological processing of their surface is extremely important.

In order to improve the wear resistance of the components the different methods of the surface treatment are used, such as mechanical-thermal treatment, metalwork and mechanical processing, surface impregnation etc. The most effective treatment techniques are those that lead to the surface strengthening and simultaneously create in the surface layers residual compressive stresses [1–4]. In this case, the resistance to the origination and propagation of crack at the surface area increase simultaneously. The strengthening also complicates the lateral movement of the surface layers, and compressive stresses prevent the crack opening and decrease the effect of tensile stresses during the component cyclic loading. The mechanical-thermal treatment combined with the doping of the surface layer is the best method that allows us to obtain such a combination of properties in the surface layers of alloys [4].

It is known [5–9] that good results for surface condition improving are achieved by forming the amorphous layer on the surface during doping and heat treatment. However, for massive components with a complex profile of the surface it is a complicated technological problem to obtain an amorphous state on entire surface.

In the present paper we describe the technology that has significantly improved the durability and corrosion characteristics of industrial pipes (such as corrosion resistance, welding, heat resistantance). This technology fits well with the existing industrial pipe manufacturing process and allows, with a small cost, to significantly increase the operational and resource characteristics of pipes and other structures made of ferrous metals based on iron alloy.

The service life of heating surfaces of the lower radiant section (LRS) of boilers with the power of 300MW and higher manufactured of steel pipes manufactured by DIN: 1.7715 (WNr: 14MoV6-3, BS: 1503-660-440) is limited to 20 thousand hours, in the area of burners and on the upper slit of the clipping tooth – 8-12 thousand hours (composition of the steel are specified in table 1). The main reason for the pipes cracking is their low corrosion resistance in combustion products of organic fuel.

Table 1. – Composition of the steel

| C, % | Si, % | Mn, % | Ni, % | S, % | P, % | Cr, % | Mo, % | V, % | Cu, % | Fe, % |
|---|---|---|---|---|---|---|---|---|---|---|
| 0.08–0.15 | 0.17–0.37 | 0.4–0.7 | <0.3 | <0.025 | <0.03 | 0.9–1.2 | 0.25–0.35 | 0.15–0.3 | <0.2 | ~96 |

One of the effective ways to increase the corrosion resistance of pipes is the application of protective metal coatings on their surface. The most significant increase in the corrosion resistance of pipes manufactured of this steel is achieved by diffusion chromizing. High corrosion resistance of chromium coatings in synthetic slags, whose composition simulates the combustion products of organic fuels, is also confirmed [10].

However, there is no information in the literature on the use in the pipes heat power engineering for the manufacturing the pipes for heating surfaces with protective chromium coatings.

Probably, this is due to the insufficient technological properties of electrodeposited chromium coatings, and in the case of diffusion chromium plating – due to the negative effect of the chromium process on the structure and properties of the parent steel.

In this connection, the aim of current work is the technology's comparison of pipes manufacturing with a protective diffusion coating of chromium, the structure and properties of the parent steel of which would meet the requirements of the heating engineering.

Three technological schemes of pipes manufacturing with a protective diffusion chromium coating have been studied. The first one is from chrome billets by hot moulding to the final size or with the production of tube stocks and subsequent cold rolling. The second one is the high-speed diffusion chrome plating of pipes with the finite size and their subsequent heat treatment. The third one is the ion-plasma diffusion coating at the atmospheric pressure condition.

In the first case, a significant deformation of the metal at pipes production from chromium billets allows complete eliminating the effect of the high-temperature heating during chrome plating. For diffusion chrome plating of billets, the vacuum resistance furnaces of the types OKV 554AM, SLV16-128-16/14 and others with similar design can be used. The chromed pipes from billets can be obtained by pressing, for example, on the presses of Zorya-Mashproekt Co, Interpipe Nico Tube, LLC "NZEST".

In the second case, when the chromium processing does not exceed 3-5 minutes, the effect of high-temperature heating is completely eliminated by subsequent heat treatment. However, this requires the vacuum furnaces of a special design.

In the third case, for the production of diffusion chrome plating the bulky expensive equipment is not required. Such equipment gives the mobility and possibility of coating under repair conditions.

The conclusion about the corrosion resistance of chrome-plated pipes can be obtained only as a result of their testing on the thermal power boilers during the burning of various types of organic fuel. The installation of pipes in the areas with the highest boilers heating surfaces damaging has been made at the Trypil'ska CHP. The Trypil'ska CHP has six power units with the power of 300MW operating on pulverized coal and gas-oil fuels.

## 2. Experiment and Discussion

### 2.1 Hot-Pressed Chrome-Plated Pipes Manufacturing

To produce the hot-pressed pipes the temporary technical specifications were developed by UkrNIIspetsstal and VNITI. According to them, the pipes manufacturing from chrome-plated welded billets of 150x270 mm diameter had to be done by pressing on the size of 32x6 mm.

The steel billets of 150x270 mm and 160x280 mm diameters, were chromium plated in the vacuum resistance furnace SLV 16-128-16/14.5x64 in the FERRO ALLOY PLANT PJSC (Zaporozhye). The temperature of the chromium plating was 1400ºC, the time of isothermal aging was 20 hours. The depth of the chromium plating was 1.7-2.8 mm depending on the level of carbon in the billets. The macrotemplate of the chrome plated billet is shown in Fig. 1.

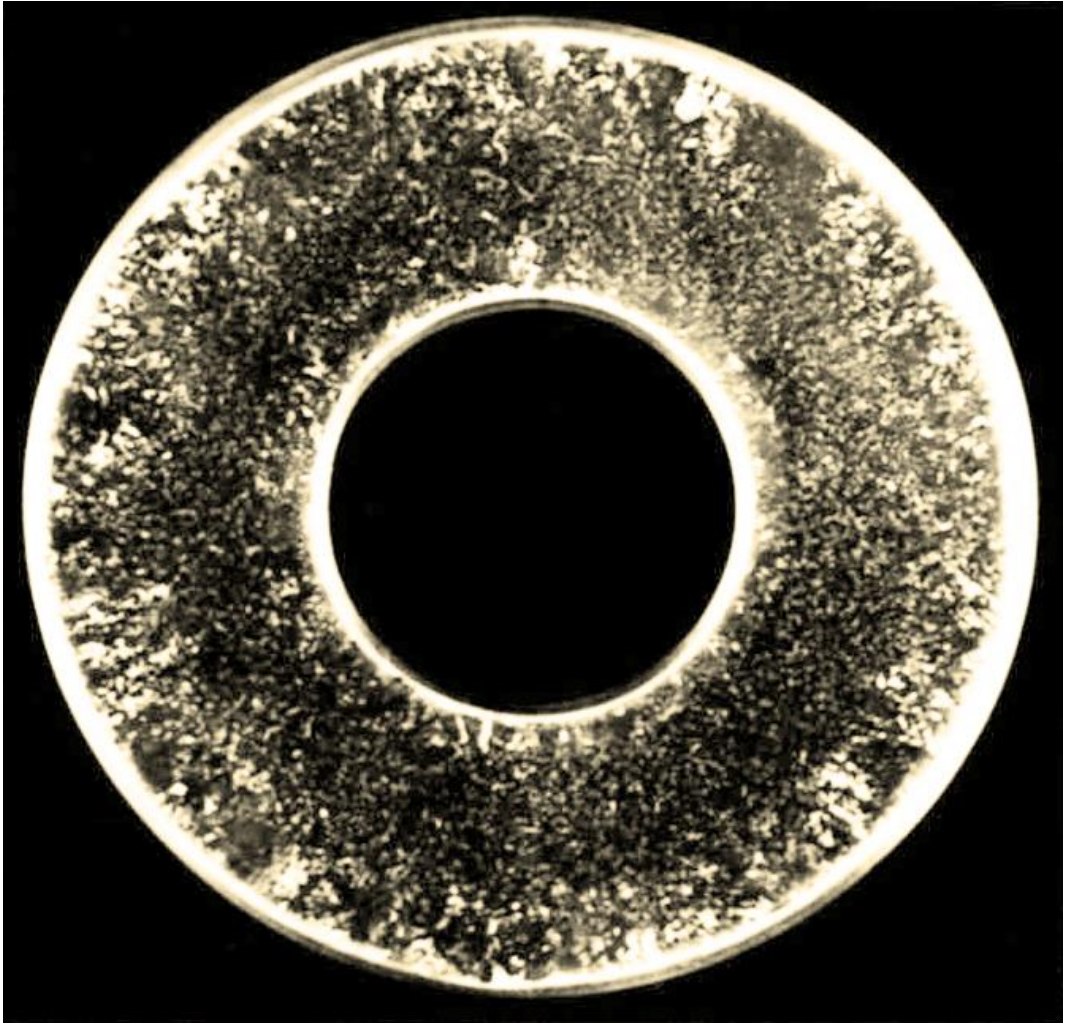

Fig. 1. The macrotemplate of the chrome plated billet.

As a depth of chromium plating the surface layer which is not etched in the 4% alcohol solution of $HNO_3$ was taken. The concentration of chromium on the surface was 60-70 %.

During the hot-pressed pipes manufacturing at the VNITI experimental plant, the dimensions of the billets and the heating technique before pressing were specified. Also, the methods for chemical treatment of pipes in order to remove lubricant after pressing and boiler scale after heat treatment (including sandblasting of the inner surface) were developed.

In order to determine the rational method of metal heating before pressing, three heating assemblies were used: horizontal and vertical inductors of high-frequency furnace OKV-665 and the chamber silicon carbide furnace OKV-210A.

The heating of the metal was carried out to the previously established temperature of 1210-1215 ºC.

The analysis showed that the worst quality of the outer surface of pipes is obtained after pressing the billets heated in a high-frequency furnace with a horizontal inductor. This is due to scoring on the surface formed when the billet moves during heating on the guide plates.

The formed scuffs in the process of molding grows into scabs and cavities.

For pipes pressed from billets heated in a vertical inductor or chamber furnace, the surface quality was better than that one in pressed billets heated in a horizontal inductor.

The difference in the metal heating time in a chamber furnace or in the inductor did not affect on the pipes quality. In terms of productivity, the induction heating should be preferred.

Due to the fact that the steam superheating pipes were manufactured by the cold deformation technique, it was advisable to produce chrome-plated pipes in the same way.

The cold deformation was carried out by single-pass and two-pass rolling with 1 hour intermediate heating at the temperature of 740ºC. For the intermediate heat treatment, it was chosen the temperature which, according to VNITI, provides a minimum strain resistance of the steel and the best resistance to fragile breakings. It was steel with 25% of chromium.

The pipes of a final size were heat treated. This treatment included the normalization at the temperature of 950-980ºC and tempering in the air at 720-750º C for 1 hour.

Under the cold strain, the pipes become hardened. In addition, the degree of hardening is greater, the greater the degree of deformation. The grains of the basis metal are stretched along the direction of deformation. Normalization, followed by tempering, results in a uniaxial grains structure.

In the case of the chrome pipes, after normalization and subsequent tempering, bainite was observed in the structure in more quantities than in the hot-pressed tubes of the same size (32×6). Apparently, this is connected with a cold strain that reduces the critical points during subsequent heating. However, this did not affect the mechanical properties of the pipes.

After heat treatment, the hardening of the pipes is completely removed, the initial properties of the pressed pipe are restored, and the yield point slightly increases. The latter is related with the age-hardening of steel accompanied by the carbides precipitation and fragmentation of the solid solution.

The mechanical properties of the pipes after the heat treatment are above the specifications and correspond to the level of hot-pressed pipes.

The analysis of the macro- and microstructure of the chrome-plated pipes after pressing, cold deformation, subsequent heat treatment shows that the chromium layer decreases during the production of pipes but remains in all cases, there is a transition zone behind it and the basis metal behind the transition zone. On the pipes obtained by pressing and cold deformation, the microhardness of the protective layer decreases by the depth from 200 to 134 $kgf/mm^2$ and from 200 to 114 $kgf/mm^2$, respectively.

The Vickers hardness of the basis metal is 200-230 $kg/mm^2$. Thus, the method of pipes producing does not affect their hardness after heat treatment.

## 2.2 Production of the developed batch of chromated pipes with diameter of 32 and 42 mm at the UMT-4 installation

At present, the pipes with chromium protective diffusion coating are not produced in our country. This is explained by the lack of high-performance technology of diffusion chrome plating, as well as by significant material softening due to prolonged exposure at high temperature.

Thus, for example, when chromizing in a backfill in a vacuum, the total duration of the process for obtaining a layer with the thickness of 0.2 mm is 25-50 hours, at the temperature of 1150°C, the strength properties are reduced in 1.5-2 times.

Under contact gas chrome plating, the duration of the process is 35-56 hours, exposure is 25 hours, at the temperature of 950-970°C. The process of pipes chrome plating can be reduced to 2-2.5 hours with non-contact gas chrome plating by the container with pipes purging with chrome chloride. However, the authors note that in this case there is considerable unevenness of the coating along the tube length.

The above circumstances did not allow the use of diffusion chromium processes to protect the pipes if high strength requirements are imposed on them.

Taking into account the exposed, in UkrNIIspetsstal the pilot plant (UMT-4) was designed and manufactured to study the continuous chrome plating technology for ready pipes, which allows to apply chromium diffusion and condensate coatings to pipes with a diameter of 25-50mm and a length of up to 6 m. The general view of the plant is shown in Fig. 2.

The technique developed by UkrNIIspetsstal together with other organizations for the production of pipes from chrome-plated billets makes it possible to obtain pipes that are not inferior in mechanical properties to pipes without coating. The developed batch of hot-pressed chrome-plated pipes after a detailed inspection was mounted for testing at 300MW blocks of the Trypil'ska CHP.

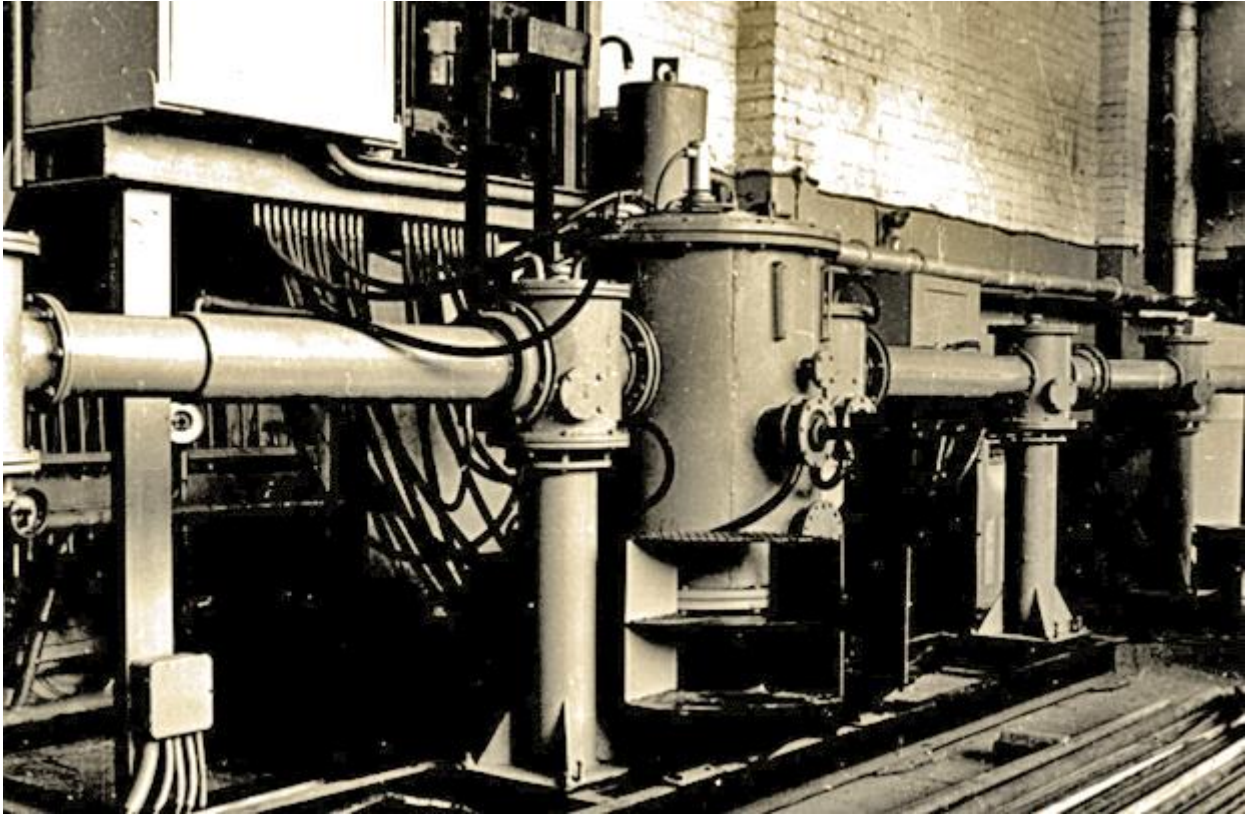

Fig. 2. – UMT-4 installation

## 2.3 Pipes chromium plating

The developed batch of tubes for diffusion chrome plating was produced at the Nikopol Yuzhnotrubnyi Plant according to technical requirments. The dimensions of pipes are 42 and 32 mm, the wall thickness is 6 mm, length is 5-9 m.

According to the agreement with UTP, the pipes for chrome plating were supplied without surface lubrication. Inspection of the pipes showed that the external surface contains a significant number of defects in the form of concavities and burrs with traces of oxide scale, the nature and size of which are allowed by the technical requirements. The presence of such defects hinders the quality of chromium plating. Therefore, the surface of the pipe before chrome plating was chipped manually or by means of grinding machine of the calibration operating department of the Dneprospetsstal plant.

The metallographic examination of the pipes showed that there is a decarbonizated metal layer from the outer and inner surfaces, the value of which varies from 0 to 200-800 μm. The pipes were chrome plated in the through-type vacuum unit UMT-4 for 3 pcs. simultaneously.

The ferrochrome with composition specified in Table 2 was used as a metallizer.

Table 2. – Composition of ferrochrome

| Cr, % | C, % | Si, % | P, % | S, % | Al, % |
|---|---|---|---|---|---|
| 65 | 0.1 | 0.5 | 0.02 | 0.02 | 0.2 |

## 3. Metallographic studies and mechanical tests of the chromium-plated pipes

After chrome plating, a sample was cut from both ends of each third pipe to control the depth of the diffusion layer and the chromium content on the surface. Selectively, on some pipes, the character of the distribution of chromium and the microhardness along the depth of the diffusion layer were analyzed. The depth of the chromium plating was controlled on the cross-sections. For the depth of the diffusion layer of chromium, a white metal layer on the surface that is not etched in the 4% solution of nitric acid was adopted.

The measurements have shown that the depth of the diffusion layer of chromium on the pipes of the tested batches varies within the limits of 0.1-0.2 mm and is practically independent on the fluctuations in the carbon content within the steel grade. This, apparently, explained by the presence of a decarbonized layer on the surface of the original pipes.

The structure of the chromium diffusion layer, γ, is solid solution of chromium in iron with a smoothly varying concentration of chromium and microhardness over the depth of the layer (Fig. 3).

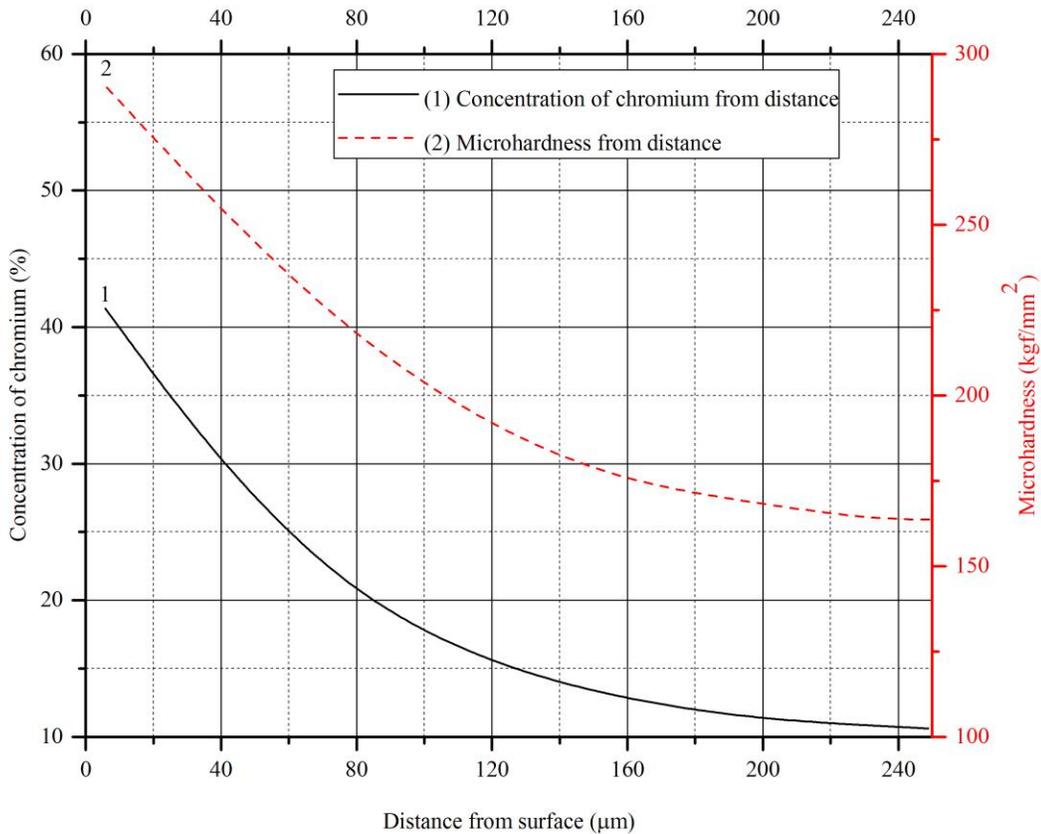

Fig. 3. – Change in the concentration of chromium (curve 1) and microhardness (curve 2) in the depth of the protective layer on the pipes.

Two characteristic curves for the distribution of chromium and microhardness in the depth of the layer with an inflection near and without the surface are noted. In the first case, there is a sputtered layer of ferrochrome on the surface, in the second one – the layer is absent.

The sputtering part of the layer appears after etching in the Marbley reagent and has a fine-grained columnar structure. Usually the deposited layer is present approximately in 0.6-0.7 of the pipe perimeter. It appears in the initial period of chromium on the part of the pipe that faces to the metallizer.

The formation of the sputtered layer is explained by the prevalence of the rate of evaporation of the metal over the diffusion rate due to weak heating of the pipe at the moment of its appearance above the metallizer. On the part of the pipe that has time to warm up, only the diffusion layer is formed (Fig. 4a). The concentration of chromium on the surface of the pipes is 35-50%. Smaller values refer to the diffusion part of the layer, large ones – to sputtered. The fact that the chromium concentration on the sputtered part is less than in ferrochrome, and also that the concentration gradient in the deposited layer is observed, indicates its formation in the initial chromium period. The boundary of the transition from the deposited part of the layer and the diffusion layer depends on the quality of the preparation of the surface for chromium plating. With careful preparation, it is practically not detected during etching of thin sections in nitric acid.

In other cases, the interface is visible in the form of a chain of pores along the boundary (Fig. 4b). Investigations of diffusion coatings with a sputtered part of the layer on the surface show that just in this part a considerable number of pores, carbonitride precipitates and sigma phases are observed. The presence of brittle structural components and pores in the sprayed part of the layer makes it undesirable.

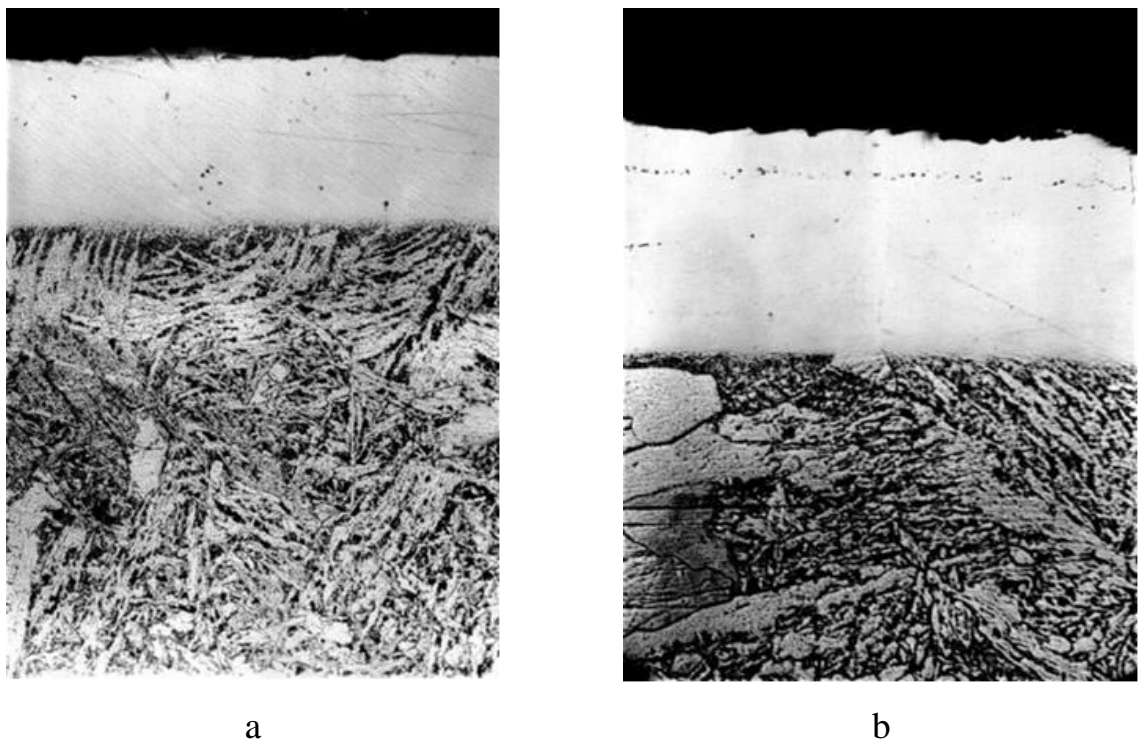

<p style="text-align:center;">a                b</p>

Fig. 4. – Microstructure of chrome plated layer on pipes, UMT-4, ×200 magnification: a) without a sputtered layer; b) with a sputtered layer.

Only the diffusion layer can be formed on the surface of the pipes either by raising the surface temperature of the pipes in the initial chromium period or by reducing the intensity of evaporation of ferrochromium.

Diffusion chrome plating of pipes at the industrial plant UMTP-11 excluded the formation of sputtered layers and allowed to regulate the concentration of chromium on the surface in any limits.

As studies have shown, the transition part of the diffusion layer ($\alpha \rightarrow \gamma$ transition in the chromium process), as a rule, does not contain characteristic for the chromium-plating steels interlayer of a metal with the pearlite structure. The microhardness curve in this region has a smooth character. This feature is explained by the presence of a significant decarburized layer on the surface, as well as by the short duration (3-5 min) of diffusion chrome plating. As a result, carbon redistribution processes do not occur in the diffusion layer.

The investigations of the structure of the basis metal indicate that, as a result of heating during the chromium plating, a coarse-grained structure consisting of 90% of bainite with a ferrite layer along the boundaries of the former austenitic grains is formed (Fig. 5a).

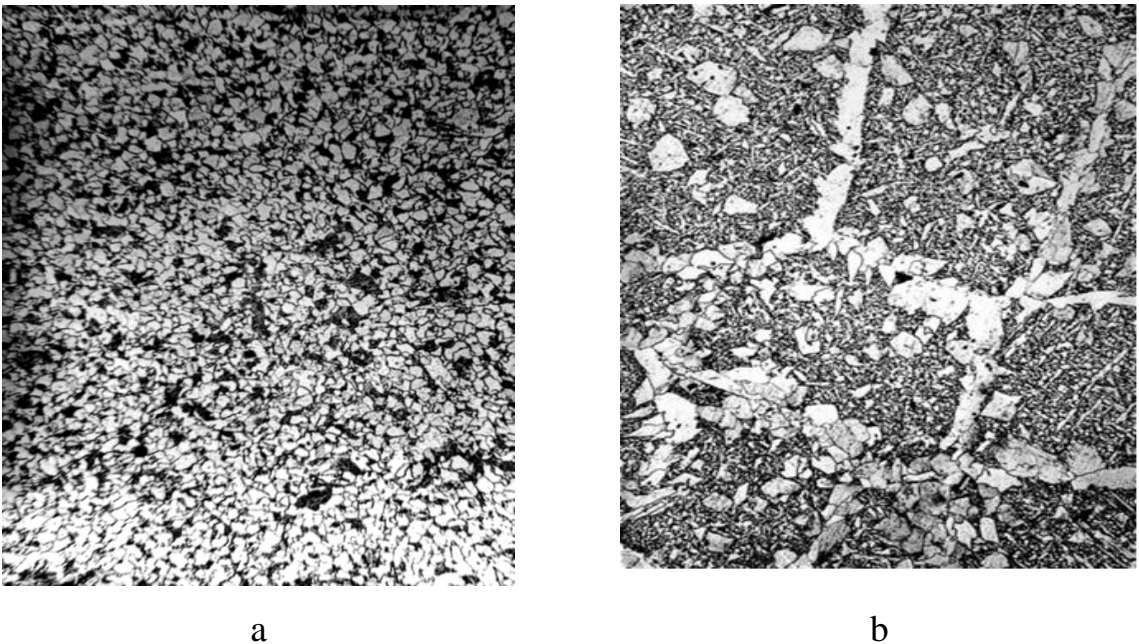

a  b

Fig. 5. – Microstructure of the basis metal on the pipes (magnification ×100): a) after chrome plating; b) after chrome plating and heat treatment.

The chromium plated pipes were heat treated according to the standard mode: normalization at 950-980°C, 30 minutes exposure; drawing-back at 720-750°C, 1 hour exposure.

The structure of the basic metal becomes a fine-grained ferrite-pearlite with a small amount of drawn-back bainite (Fig.5b).

The comparison of the microstructure of chrome and non-chrome pipes of the same melting after normalization with tempering under identical conditions shows that the chromium-plated pipe has a smaller structure and the amount of tempered bainite in it is smaller. This affected on the level of mechanical properties of pipes – the strength characteristics are lower, and the plastic ones are higher in chrome pipes than in non-chromated pipes, although they meet the technical requirements (Table 3.).

Table 3. – Mechanical characteristics of pipes

| Material | Melting number | lot | Dimensions of pipes, mm | Mechanical characteristics | | |
|---|---|---|---|---|---|---|
| | | | | $\sigma_B$ resistance to rupture | $\sigma_T$ flow limit | $\delta_5$ elongation at failure |
| Source steel | 5830  5849 | 1250  1218 | Ø42x6  ÷ | 57  54 | 45  41 | 28  26 |
| Chrome-platted steel | 5830 | 1250 | Ø42x6 | 52 | 41 | 26 |
| Technical requirements | - | - | - | 45 | 28 | 21 |

Note: Chromium-plated pipes had satisfactory flattening with the allowed technical requirements – for this steel the distance between the flattened surfaces is 24 mm and less.

# 4. Investigation of the technological properties of chromium-plated pipes

As it was noted above, the structure of the basis metal of the chrome pipes after heat treatment, irrespectively to the method of production (hot pressed or chrome plated on the UMT-4 plant), meets the requirements. The diffusion chrome plating does not also have a significant effect on the mechanical properties of pipes, slightly reducing the strength and increasing the plastic properties of the metal. Despite the high concentration of chromium in the surface diffusion layer, the latter has a sufficiently high plasticity and does not break down when testing the pipes for flattening and bending. Other requirements for chrome pipes are high corrosion resistance in combustion products of organic fuel and good weldability.

## 4.1. Corrosion resistance

The main purpose of the diffusion layer of chromium on the boiler pipes is to protect it from corrosion in the combustion products of organic fuel. The available information on the composition of the gas phase and fuel ash deposits on the surface of the pipes, obtained as a result of the analysis at temperatures different from that one of the pipes operation, makes it difficult to assess the responsibility of various elements or compounds for the course of corrosion processes.

Based on the analysis of corrosion products, there are two main types of corrosion: sulphide and the so-called "vanadium".

In sulphurous corrosion, the main cause of the destruction of the pipes surfaces is the formation of complex sulfates of $K_3Fe(SO_4)_3$; $Na_3Fe(SO_4)_3$ with a melting point about 600ºC which is close to the operating temperature of the boiler pipes.

For complex alloy steels, sulfates can also be formed with lower melting points.

In the case of vanadium corrosion, low-melting compounds are also formed with a melting point close to the melting temperature of sulfates.

The presence of the diffusion layer of chromium on the pipe surface excludes the possibility low-melting compounds formation at the temperature of the boiler pipes operation. Considering that the existing methods of corrosion resistance testing in synthetic slags cannot fully characterize the serviceability of pipes in combustion products of organic fuel, the research of pipes was carried out directly on boilers of thermal power units. The developed batch of pipes was installed in the lower radiation part of the TGMP-314 boiler.

The fuel of the boiler was gas with fuel oil.

After 21 thousand hours of the boiler operation, a partial cutting of pipes and a study of their condition were carried out. It has been established that the appearance and diameter of the chrome pipes have not changed. In ash deposits on the adjacent side and pipes, chromium was not detected by spectral analysis.

These facts indicate that there is no general corrosion of chrome plated pipes.

Metallographic examination of pipes that have worked 21 thousand hours showed that the continuous diffusion layer of chromium was preserved, but the thickness of the layer slightly increased in comparison with the initial state, that is explained by the processes of secondary diffusion of chromium into the base of the pipes.

Analyzing the results of corrosion tests of chromium-plated pipes on a TGMP-314 boiler, it can be assumed that the service life of chrome-plated pipes will be determined mainly by the kinetics of the secondary chromium diffusion process in the operation of pipes.

An approximate estimation of the surface of chrome-plated pipes destruction beginning can be made using the following equation:

$$\tau = \frac{\delta^2 tg^2\left[\frac{\pi}{2}\left(1-\frac{C_m}{C_o}\right)\right]}{4Д} \quad (1)$$

where $C_o$ and $C_m$ are the initial and final chromium concentrations on the surface of the pipes, correspondingly; $\delta$ is the depth of the diffusion layer, D is the diffusion coefficient of chromium, and $\tau$ is the time of decrease in chromium concentration on the surface of pipes from Co to Cm.

It follows from Eq. (1) that as the initial concentration of chromium on the surface and the depth of the layer increase, the time of its operation increases.

Figure 6 shows curves of the change in chromium concentration on time for different thicknesses of the diffusion layer. If in the capacity of the time of active oxidation beginning we take the time when the chromium concentration on the surface drops to 12%, then for pipes with the initial chromium concentration of 25% and the diffusion layer thickness of 0.05, 0.10 and 0.15 mm, respectively, it will be equal to 15, 60 and 120 hours.

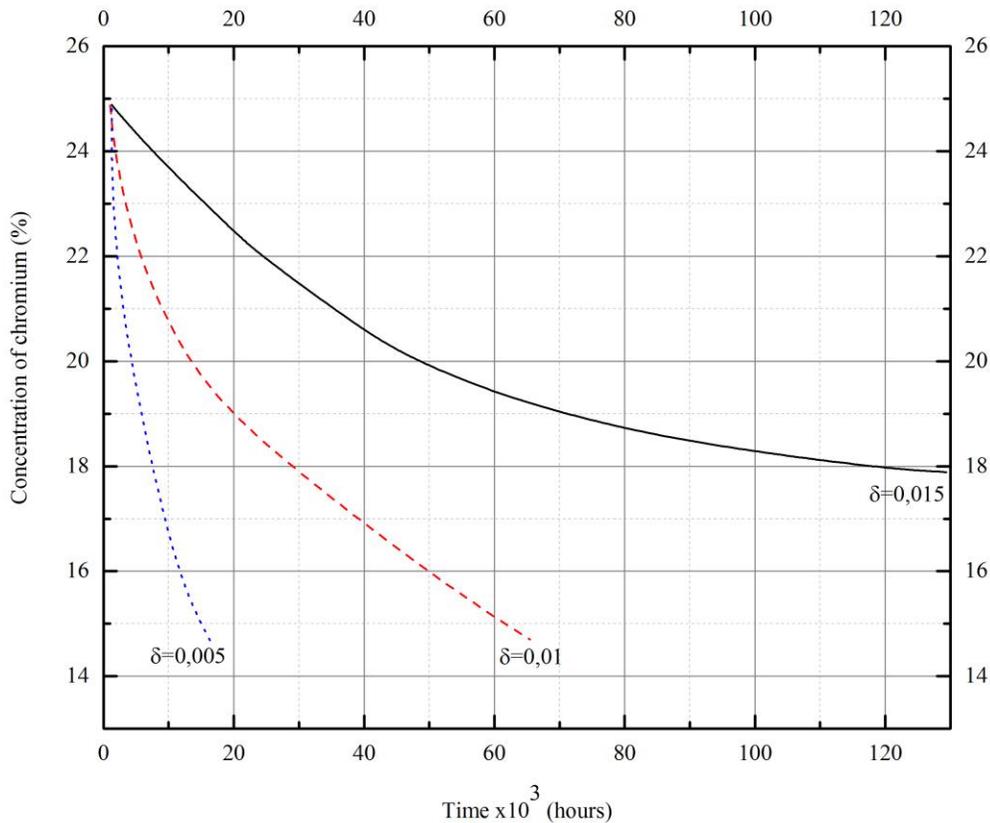

Fig. 6. Chromium concentration change in the layer during operation.

Taking into account the cyclic nature of the pipe loading, wide range of operating temperatures in the zone of lower radiant section that include the brittle transition temperature of chromium steels, probability of short-term overheating of pipes during operation, presence of welds that differs in structure and composition from the diffusion layer, other types of protective layer destruction can be expected along with general corrosion.

The investigations of the developed batch of chrome-plated pipes at Trypil'ska CHP allow to answer these and other questions related to the resistance of chrome pipes.

## 4.2 Heat resistant properties, welding

Two variants of welding of chrome plated pipes were investigated. The first one is the welding of the entire weld by ZL-39 (ESAB: OK 76.18) electrodes at butt-welding of chromium-plated pipes with non-chromated ones. The second one is welding with ZL -39 electrodes with overlapping of the top layer by ZT-15 (ESAB: OK 61.85; OK 61.80; OK 61.86) electrodes at docking of the chrome plated pipes with each other.

The welding was performed in accordance with guidance document "Welding, heat treatment and control of pipe systems of boilers and pipelines in the installation and repair of power plant equipment". Before welding, the edges of pipes were cut at the angles of 40-45 degrees. The diameters of the wire of electrodes are 2.5 mm for ZL-39 and 3 mm for ZT-15. At visual control, the welds are in satisfactory condition. The cracks, pores, shells, undercuts and other external defects were not observed. The internal defects in butt welds by ultrasonic inspection were not detected. The presence of a diffusion layer of chromium on the pipes surface did not prevent the ultrasonic inspection of welded joints. The microstructure of the metal in the weld zone and the joints is satisfactory in both cases. The damage of the diffusion layer in the near-weld zone was not observed.

## Conclusions

The diffusion chrome plating of pipes allows to increase their corrosion resistance in combustion products of organic fuel (pulverized coal and gas-oil fuel) by several times.

The durable strength of 100'000 hours for chromium-plated pipes is on the same level as for non-chrome plated pipes.

**References**

[1] G.H. Farrahi, H. Ghadbeigi, An investigation into the effect of various surface treatments on fatigue life of a tool steel, Journal of Materials Processing Technology 174 (2006) 318–324.

[2] C. Swett, Outpatient phenothiazine use and bone marrow depression. A report from the drug epidemiology unit and the Boston collaborative drug surveillance program, Archives of general psychiatry 32 (1975) 1416–1418.

[3] Z.L. Zhang, T. Bell, Structure and Corrosion Resistance of Plasma Nitrided Stainless Steel, Surface Engineering 1 (2013) 131–136.

[4] G.E. Totten, Steel heat treatment: Metallurgy and technologies / George E. Totten, editor, Taylor & Francis, Boca Raton, FL, 2007.

[5] C. Borchers, T. Al-Kassab, S. Goto, R. Kirchheim, Partially amorphous nanocomposite obtained from heavily deformed pearlitic steel, Materials Science and Engineering: A 502 (2009) 131–138.

[6] A.V. Byeli, O.V. Lobodaeva, S.K. Shykh, V.A. Kukareko, Solid-state amorphization of a tool steel by high-current-density, low-energy nitrogen ion implantation, Nuclear Instruments and Methods in Physics Research Section B: Beam Interactions with Materials and Atoms 103 (1995) 533–536.

[7] Daniel J. Branagan, Joseph V. Burch, Methods of forming steel, US6258185B1 (1999), US.

[8] K. Hashimoto, N. Kumagai, H. Yoshioka, J.H. Kim, E. Akiyama, H. Habazaki, S. Mrowec, A. Kawashima, K. Asami, Corrosion-resistant amorphous surface alloys, Corrosion Science 35 (1993) 363–370.

[9] Y.N. Petrov, V.G. Gavriljuk, H. Berns, F. Schmalt, Surface structure of stainless and Hadfield steel after impact wear, Wear 260 (2006) 687–691.

[10] Shang-Hsiu Lee, High-Temperature Corrosion Phenomena in Waste-to-Energy Boilers. Submitted in partial fulfilment of the , USA, Columbia, 2009.